# Unusual Nonlinear Optical Responses in Layered Ferroelectric Niobium Oxide Dihalides: Origin and Manipulation

Liangting Ye[1#], Wenju Zhou[2#], Dajian Huang[2], Xiao Jiang[1], Donghan Jia[2], Qiangbing Guo[3], Dequan Jiang[2], Yonggang Wang[2], Xiaoqiang Wu[4], Yang Li[1]*, Huiyang Gou[2]*, and Bing Huang[1,5]*

[1]Beijing Computational Science Research Center, Beijing 100193, China

[2]Center for High Pressure Science and Technology Advanced Research, Beijing 100193, China

[3]Department of Electrical and Computer Engineering, National University of Singapore, Singapore, Singapore

[4] School of Mechanical Engineering, Chengdu University, Chengdu 610106, China

[5] Department of Physics, Beijing Normal University, Beijing 100875, China

#These authors contributed equally: L.T.Y. and W. Z.

E-mails: liyang2020@csrc.ac.cn (Y.L.); huiyang.gou@hpstar.ac.cn (H.G.); Bing.Huang@csrc.ac.cn (B.H.)

## Abstract

**Realization of large and highly tunable second-order nonlinear optical (NLO) responses, e.g., second-harmonic generation (SHG) and bulk photovoltaic effect (BPVE), is critical for developing modern optical and optoelectronic devices. Very recently, the two-dimensional van der Waals ferroelectric NbO$X_2$ ($X$ = Cl, Br or I) are discovered to exhibit unusually large and anisotropic SHG. However, the physical origin and possible tunability of NLO responses in NbO$X_2$ remain to be unclear. In this article, we reveal that the large SHG in NbOCl$_2$ is dominated by the synergy between large transition dipole moment and band-nesting-induced large intensity of electron-hole pairs. Remarkably, the NbOCl$_2$ can exhibit dramatically different strain-dependent BPVE under different polarized light, originating from the interesting light-polarization-dependent orbital transition. Importantly, we successfully achieve a reversible ferroelectric-to-antiferroelectric**

**phase transition via controlling ambient temperature or external pressure, accompanied by the greatly tunable NLO responses. Furthermore, we discover that the evolutions of SHG and BPVE in NbO$X_2$ with variable *X* obey different rules. Our study provides a deep understanding on the novel NLO physics in NbO$X_2$ and establishes great external-field tunability for device applications.**

## Introduction

The responses of materials under the optical field are the basis of many novel physical phenomena and detection techniques. Among numerous materials, compounds without inversion symmetry can exhibit various nonlinear optical (NLO) responses. The second-harmonic generation (SHG), referred the effect that the frequency of light is doubled when it passes through a material, is a typical example of the NLO effect, which has been not only used to characterize the basic properties of materials, such as structural symmetry [1,2], magnetic ordering [3,4], and polarized domains [5], but also used to design electro-optical devices, such as frequency conversions [6,7], electro-optic modulators [8,9] and optical switches [10]. Besides the SHG, the shift current, referred the direct current induced by the change of wave functions when the electrons are excited from the valence band to the conduction band, is another important NLO phenomenon. The shift current is the main mechanism of bulk photovoltaic effect (BPVE) [11,12], which is promising to overcome the Shockley–Queisser limit of solar energy conversion [13].

Searching materials with large and greatly tunable NLO responses is long sought for various NLO device applications [8]. The advent of low-dimensional materials provides new opportunities to find candidates with large second-order NLO responses, such as transition-metal-dichalcogenide (TMD) layers [2,14] and nanotubes [15], twisted boron nitrides [16], and 2D multiferroics [17], which is more suitable for the design of nanoscale electronic and photonic devices compared with traditional bulk materials. Very recently, the layered ferroelectric (FE) material niobium oxide dihalides NbO$X_2$ ($X$ = Cl, Br, and I) has attracted plenty of interests due to their large, anisotropic, and even layer-independent SHG responses [18,19], significantly different from many known 2D NLO materials [2,20,21]. Consequently, the NbO$X_2$ thin-film can be used to design a spontaneous parametric down-conversion quantum light source with a record performance [19]. Interestingly, although holding different halides, the observed SHG strength in NbOCl$_2$ and NbOI$_2$ are quite similar [18,19]. Given the rapid progress of exciting experiments, the microscopic origin of SHG effects in NbO$X_2$ remains to be largely known, e.g., it is unclear whether the NLO effects of NbO$X_2$ are strongly coupled with their ferroelectricity. On the other hand, developing effective means to manipulate the NLO responses is highly desired for FE NbO$X_2$, which is the key step for realizing NbO$X_2$-related optical sensing, optical computing, and optical switch.

In this article, we reveal the physical origin of large NLO responses observed in NbO$X_2$ using density functional theory (DFT) based NLO calculations (see Method). We discover that the large SHG observed in NbOCl$_2$ is mainly contributed by the double-photon resonances between occupied anion $p$ orbitals and empty Nb $4d$ orbitals, originating from the synergy between the large transition dipole moment and band-nesting-induced large joint density of states (JDOS). Besides SHG, the NbOCl$_2$ can exhibit very different strain-dependent shift current responses under different polarized light, originating from the selective optical transition between different orbitals that hold anisotropic deformation potentials. Combining with the single crystal X-ray diffraction and SHG measurements, we observe that NbOCl$_2$ can experience a novel structural phase transition from the FE phase to the antiferroelectric (AFE) phase as a function of temperature or external pressure, accompanied by greatly tunable material polarization and NLO effects. Finally, we discover that the SHG (shift current) response in NbO$X_2$ is insensitive (sensitive) to the different $X$.

## Results

**Structural and Electronic Properties of NbO$X_2$.** Niobium oxide dihalides NbO$X_2$ are a series of vdW layered materials [22-24]. As shown in **Fig. 1a**, the individual vdW layers are ABC stacked along *z* direction, crystallized to a monoclinic structure with the $C_2$ space group. Within each vdW layer, Nb atoms are inside the edge sharing distorted octahedra chains which mutually connect via O atoms along *b* direction. Compared with other transition-metal oxide dihalides [25,26], the combined displacements of Nb atoms along *c*, *b* directions create a unique crystal structure of NbO$X_2$: (i) the Nb atoms are dimerized along *c* direction (i.e., $d_l>d_s$), resulting in a noticeable Peierls distortion, accompanied by a metal-to-insulator transition (see Supplemental **Fig. S1**); (ii) the off-center displacements of Nb atoms along *b* direction causes the separation of positive and negative charge centers, further forming spontaneous polarization (***P***) and breaking the inversion symmetry (**Fig. 1b**, *Left* panel). Interestingly, whether the appearance of FE phase could enhance the NLO responses is still an open question [27,28]. In principle, the shift of several adjacent Nb atoms to the opposite directions may create other metastable phases, i.e., AFE phase (**Fig. 1b**, *Right* panel), although the possible AFE phases have never been observed in the experiments. In contrast to the FE phase, the inversion symmetry of AFE NbO$X_2$ could forbid all the second-order NLO responses.

For NbO$X_2$ family, the extremely weak interlayer coupling is a unique property that is different from many other 2D materials [19], reflected by the layer-insensitive electronic band structures (see Supplemental **Fig. S2**). Taking monolayer NbOCl$_2$ as an example, **Fig. 1c** lists the calculated basic parameters in the ground-state FE phase. As one can see, the lattice constant *c* increases obviously with the change of *X* from Cl to I, expanding the space of O$_2X_4$ octahedra and enhancing the Peierls distortion (i.e., increase of $d_l/d_s$). In contrast, the point-sharing connection of the O$_2X_4$ octahedra causes the lattice constant *b* insensitive to *X*. Moreover, the increase of atomic radius with the change of *X* from Cl to I can also increase O-Nb-*X* bond angle, weakening the electrical dipole moment for O$_2X_4$ octahedra, which is consistent with the decreased polarization ***P***. The rather different changes of the lattice constants *c* and *b* with varying halogen clearly suggest the remarkable crystal anisotropy of NbO$X_2$. In the following, we mainly focus on the NbOCl$_2$, and the results of other isostructural NbO$X_2$ will be discussed at the end of this paper.

**Figure 1d** shows the HSE-calculated band structure of monolayer NbOCl$_2$, which exhibits a feature of indirect bandgap ~1.98 eV [19]. According to the crystal field theory and Jahn-Teller distortion [29], the Nb $d$ orbitals will split into three nondegenerate subgroups (i.e., $d_z^2$, $d_{xz}$, $d_{x^2-y^2}$) and a double-degenerate subgroup (i.e., $d_{xy}$ and $d_{yz}$). As the Nb atom has the valence electronic configuration of $4d^45s^1$, there is only one unpaired $4d$ electron for each cation Nb$^{+4}$ in NbOCl$_2$, half filling the Nb $4d_z^2$ orbitals (**Fig. 1e**). The dimerization of Nb atoms along $c$ direction, in the form of Peierls distortion, further splits the Nb $4d_z^2$ orbitals in energy. This Peierls distortion can create an isolated flat bonding-like Nb $4d_z^2$ band around the Fermi level, forming the top of the valence band (VB) with a large density of states in NbOCl$_2$. Meanwhile, the empty antibonding-like Nb $4d_z^2$ orbitals are pushed to ~ 1 eV above the bottom of the conduction band (CB). Noticeably, the low-energy O $2p$ and Cl $3p$ orbitals mainly contribute to the VB around 2 eV below the Fermi level, well separating from the flat band.

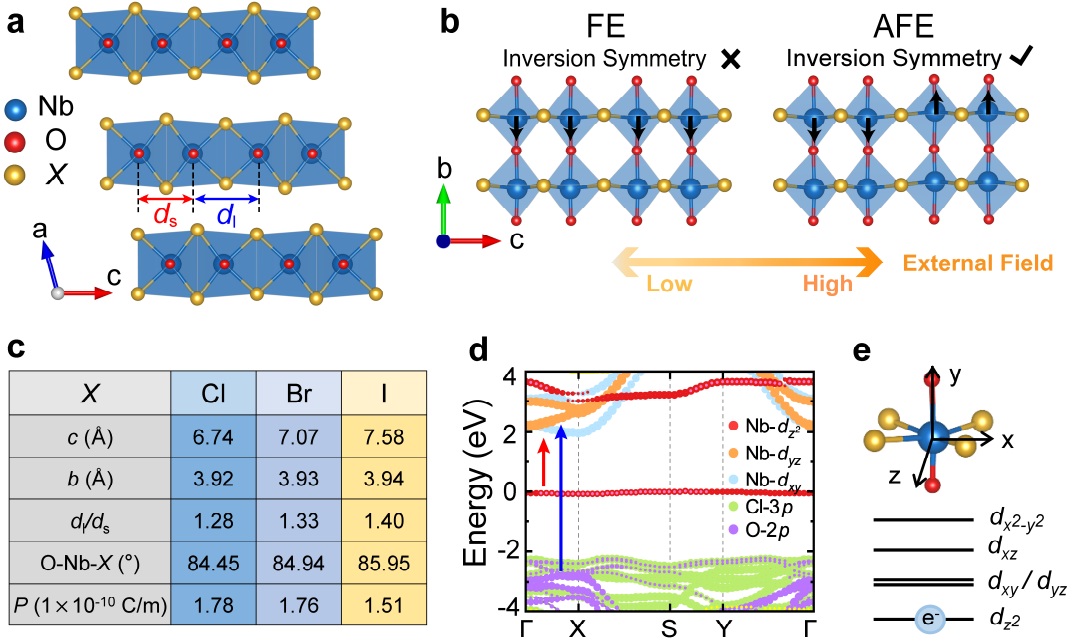

**Fig. 1 Crystal and Electronic Properties of NbO$X_2$. a** Side view of the crystal structure of NbO$X_2$. Here, the blue, red, and yellow balls represent Nb, O, and $X$ atoms, respectively. $d_l$ and $d_s$ are the distances between the Nb atomic pairs. **b** Top view of the monolayer structure for NbO$X_2$ in FE (*Left* panel) and AFE (*Right* panel) phases, respectively. FE-to-AFE phase transition may be realized via applying external fields. **c** Summarization of the calculated parameters of monolayer NbO$X_2$ in the ground-state FE phase. Here, $c$ and $b$ represent lattice constants, $P$ indicates polarization, O-Nb-

$X$ means bond angle, $d_l$ and $d_s$ are distances shown in **a**. **d** HSE-calculated band structure of monolayer NbOCl$_2$ with orbital projection. Here, the red and blue arrows represent two important optical transitions between Nb $d$ – Nb $d$ orbitals and between Nb $d$ – Cl $p$ orbitals, respectively. **e** Local geometries of NbOCl$_2$ (*upper* panel) and the schematic energy splitting of Nb-$d$ orbitals (*bottom* panel). Here, the blue ball means an unpaired electron.

**Unusual SHG Responses in NbOCl$_2$.** **Fig. 2a** shows the typical SHG process, where the frequency $\omega$ of light is doubled when it passes through a NLO material. This SHG effect can be described as [30]:

$$P^a(2\omega) = \epsilon_0 \chi^{(2)}_{abc} E^b(\omega) E^c(\omega), \quad (1)$$

where $\boldsymbol{P}(2\omega)$ represents the polarization, $\chi^{(2)}$ is known as SHG susceptibility, $\boldsymbol{E}(\omega)$ indicates the electric field of the incident light, $\epsilon_0$ is the dielectric constant of vacuum, and $a, b, c$ are the indices in Cartesian coordinates. For the non-centrosymmetric materials, the number of independent elements of the three-order tensor $\chi^{(2)}$ is governed by the direct product $\Gamma_P \otimes \Gamma_{EE}$ [31]. For monolayer NbOCl$_2$ with the $C_{2v}$ point group, the representations $\Gamma_P$ and $\Gamma_E$ can be divided into three irreducible representations: $\Gamma_P = \Gamma_E = A_1 + B_1 + B_2$. For linear susceptibility, we have $\Gamma_P \otimes \Gamma_E = 3A_1 + 2A_2 + 2B_1 + 2B_2$, leading to three independent nonzero components. The direct product $\Gamma_P \otimes \Gamma_E$ can be further divided into a symmetric part $\Gamma^s = 3A_1 + A_2 + B_1 + B_2$ and an anti-symmetric part $\Gamma^a = A_2 + B_1 + B_2$ [31]. Hence for the symmetric third-rank tensor, we can write $\Gamma_P \otimes \Gamma^s = 5A_1 + 3A_2 + 5B_1 + 5B_2$, suggesting only five independent nonzero elements in the $\chi^{(2)}$ tensor for monolayer NbOCl$_2$.

Indeed, the calculated SHG susceptibilities have five nonzero components, i.e., $\chi^{(2)}_{xxy}$, $\chi^{(2)}_{yxx}$, $\chi^{(2)}_{yyy}$, $\chi^{(2)}_{yzz}$, and $\chi^{(2)}_{zyz}$, as shown in **Fig. 2b**. Among them, the components along the non-polarized directions, i.e., $\chi^{(2)}_{xxy}$ and $\chi^{(2)}_{zyz}$, are small in a large photon energy range. On the contrary, the components along the polarized directions display noticeable values, a feature clearly demonstrating the unique role of ferroelectricity in generating anisotropic SHG responses in NbOCl$_2$. The component $\chi^{(2)}_{yxx}$ shows great values within the range of 2~4 eV, with the maximum of ~120 pm/V at $\omega = 2.2$ eV. The component $\chi^{(2)}_{yzz}$ has the similar maximum of ~110 pm/V at the higher position of $\omega = 4.8$ eV. The most important component is $\chi^{(2)}_{yyy}$, which corresponds to the SHG response induced by the *y*-polarized incident light. Interestingly, the $\chi^{(2)}_{yyy}$ displays the large value over a wide energy region and reaches its maximum of ~160 pm/V at $\omega = 2.4$ eV, denoted as peak B. The strength of peak B is much larger than the peaks of $|\chi^{(2)}|$ in BN sheet (~20 pm/V) [2] and LiNbO$_3$ (~50 pm/V) [32]. Besides peak B, there is also a small peak appeared at $\omega = 1.14$ eV in $\chi^{(2)}_{yyy}$, denoted as peak A. To analyze the SHG effect of NbOCl$_2$, we also calculate the angle-resolved SHG polarization of monolayer NbOCl$_2$ under $\omega = 2.4$ eV, as shown in **Fig. 2c**. With the azimuth angle ($\theta$) [marked in **Fig. 2a**] changes, the SHG response exhibits a significant anisotropy and shows a two-fold rotational symmetry, consistent with the symmetry of crystal structure. Again, this strong anisotropic feature is absent in many existing 2D materials [20,21].

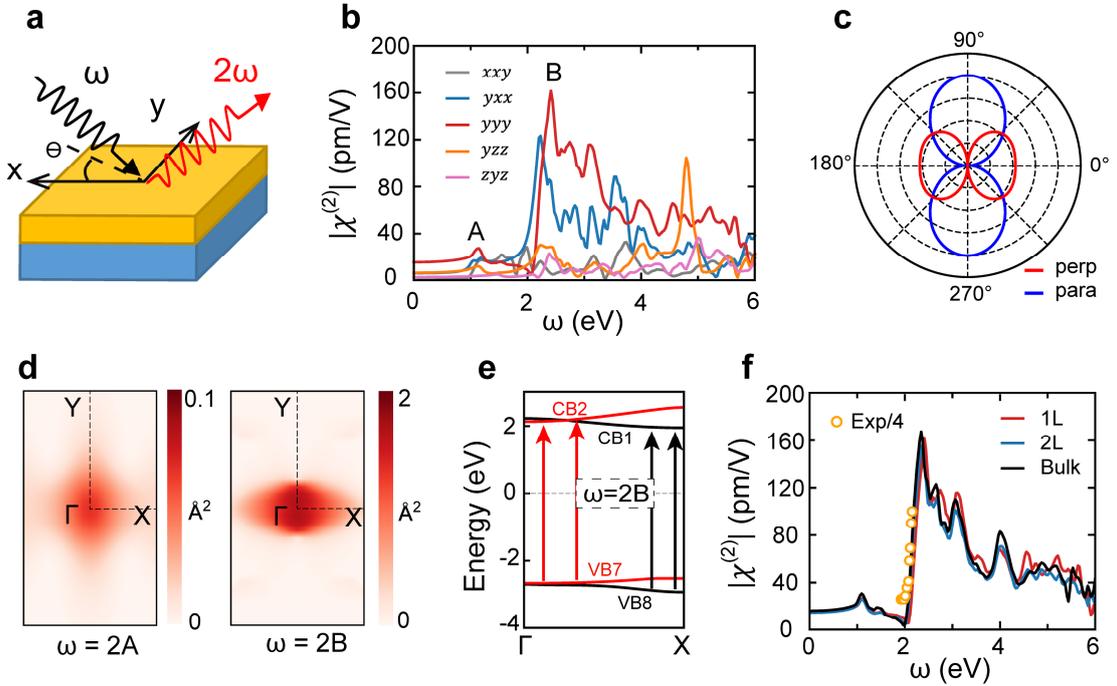

**Fig. 2 SHG Responses in NbOCl$_2$. a** Schematic diagram of the SHG process. **b** Calculated nonlinear susceptibility $|\chi^{(2)}_{abc}|$ as a function of incident photon energy. **c** Angle-resolved SHG polarization. Here, the photon energy of the incident light is set to be 2.4 eV, and the red (blue) curve represents the response with the direction perpendicular (parallel) to the polarization of incident light. **d** *k*-resolved optical absorption strength near SHG peaks, representing *d-d* transition (*Left* panel) and *p-d* transition (*Right* panel) over the first Brillouin zone. **e** Schematic diagram of the strong optical transition caused by the band nesting along Γ-X line. Here, VB$_n$ (CB$_n$) labels the $n^{th}$ valance (conduction) band starting from the top valence (bottom conduction) band. **f** SHG response for monolayer, bilayer, and bulk NbOCl$_2$. Experimental data [19] is plotted with the orange circles, with a blue shift of ~0.6 eV for comparison.

We focus on the understanding of the origin of peaks A and B in $\chi^{(2)}_{yyy}$, and similar peaks exist in other nonzero $\chi^{(2)}$ components. In general, the SHG is closely related to the transition dipole moment $r^a_{nm}$ and JDOS [33,34] (see Method). Interestingly, for the $r^a_{nm}$, the calculated results under *y*-polarized incident light exhibit different origins for peak A and peak B. As shown in **Fig. 1d** (red arrow), the peak A is contributed by the double-photon resonances between the flat VB dominated by the Nb $d_z^2$ orbital and bottom of CB dominated by the Nb $d_{xy}$ orbitals, i.e., the $d \rightarrow d$ transition, while the peak B is mainly contributed by the double-photon resonances between the VB below ~ -2 eV dominated by Cl 3*p* and O 2*p* orbitals and bottom of CB, i.e., the $p \rightarrow d$ transition (blue arrow in **Fig. 1d**). **Figure 2d** shows the *k*-resolved optical absorption strength, i.e., $r^y_{nm} r^y_{mn}$, which can characterize the strength of interaction between the system and the electromagnetic wave quantitatively. The absorption strength of peak B is strong, resulting in a relatively large SHG value. Meanwhile, as shown in **Fig. 2e**, the double-photon resonances dominated peak B are mainly caused by the transition between 7$^{th}$ (8$^{th}$) VB and 2$^{th}$ (1$^{th}$) CB. These bands for optical transition are almost parallel to each other along the $\Gamma - X$ path, resulting in a significant band nesting effect and creating the singularities in the JDOS. Therefore, the singularities in JDOS can further enhance the SHG response at peak B [35]. On the other hand, the absorption strength of peak A is much weaker than that of peak B (**Fig. 2d**), consistent with the weak SHG response.

Furthermore, we have calculated the $\chi^{(2)}_{yyy}$ spectra of NbOCl$_2$ with different layer thicknesses. As shown in **Fig. 2f**, the SHG strength of NbOCl$_2$ is almost independent of the sample thickness, different from other vdW layered NLO compounds such as TMDs [2,36] and SnS (Se) [37], where the SHG strength will gradually decrease with increasing the sample thickness. In addition, we have compared the available experimental SHG data with our calculated one. As shown in **Fig. 2f**, our calculated spectrum has a similar curvature to the experimental one [19], indicating that peak B may account for the observed large SHG in NbOCl$_2$ in experiment. We emphasize that the exclusion of exciton effect [38] could result in avoidable differences between our calculations and experimental data, including the exact peak position and peak intensity in the SHG spectra, which deserves future investigation. Meanwhile, the values of smearing factor ($\eta$) and electronic relaxation time ($\tau$) may also affect exact values of SHG [35,39].

**Shift Current in NbOCl$_2$**. Besides the large SHG response, the non-centrosymmetric structure also allows NbOCl$_2$ to generate nonzero shift current density $J$ under external electric fields [30]:

$$J_{sc}^a = 2\sigma^{abc}(0;\omega,-\omega)E^b(\omega)E^c(-\omega), \quad (2)$$

where $\sigma$ is the shift current susceptibility tensor. Similar to SHG, there are only five independent susceptibility components in monolayer NbOCl$_2$ owing to the restriction of crystal symmetry (see Supplemental **Fig. S3**). **Figure 3a** shows two in-plane $\sigma$ with the most remarkable values. As one can see, the component $\sigma^{yxx}$ caused by the $x$-polarized incident light reaches the maximum when $\omega \approx 3.5$ eV and reverses its sign when $\omega > 4.0$ eV. However, the component $\sigma^{yyy}$ induced by the $y$-polarized incident light has large values over a relatively wide range of $4.3 < \omega < 5.0$ eV.

The $\sigma$ can be considered as an integration of shift vector weighted by optical transition rate over the entire Brillouin zone [27,40]. To illustrate the mechanism of $\sigma$, we have calculated the integrated shift vector $e\bar{R}^{ab}$ and the linear optical absorption spectrum $\varepsilon_2^{ab}$ (see Method). Because the phase of transition dipole and Berry connection appeared in shift vector contain the information about the change of the charge center in real space during excitation, the $e\bar{R}^{ab}$ usually can determine the main features of $\sigma$. Indeed, as shown in **Figs. 3a** and **3b**, the spectrum of $e\bar{R}^{yy}$ has a similar shape with that of $\sigma^{yyy}$, and the amplitudes of $e\bar{R}^{yy}$ with positive and negative values are also consistent with that of $\sigma^{yyy}$. On the other hand, the $\sigma^{yxx}$ seems to be mainly determined by the $\varepsilon_2^{xx}$. Especially, the large peak in $\varepsilon_2^{xx}$ is consistent with the peak A of $\sigma^{yxx}$, as shown in **Figs. 3a** and **3c**.

Furthermore, we evaluate the change of $\sigma$ under in-plane strain, and the corresponding results are shown in **Figs. 3d** and **3e**. Due to the significant anisotropy of NbOCl$_2$, the major source of optical transition is dependent on the polarization of the incident light. For the $x$-polarized light, the optical transition is dominated by the coupling between two out-of-plane Nb $d_z^2$ states, where the dispersion and energy gap between these two states are less sensitive to the in-plane strain (see Supplemental **Fig. S4**). Therefore, the optical transition rate remains almost unchanged under in-plane uniaxial strain, leading to the insensitive responses of $\sigma^{yxx}$ to external strain (see **Fig. 3d**). For the *y*-polarized light, the optical transition is mainly contributed by the coupling between anion *p* states and Nb *d* states. The energy gap between *p* and *d* states retains under strain along *x* direction, but increases when the strain along *y* direction changes from -2 % to 2 % (see Supplemental **Fig. S4**). As a result, the optical transition rate almost has no change under strain along *x* direction but gradually decreases under strain along *y* direction, in line with the calculated results that the $\sigma^{yyy}$ is insensitive to the strain along *x* direction but sensitive to the strain along *y* direction (see **Fig. 3e**). This unique strain-dependent shift current response, originating from the selective optical transition between different orbitals that hold anisotropic deformation potentials to external strain [41], is quite different from other 2D material systems [42].

Besides, it is noteworthy that the $\sigma$ of NbOCl$_2$ is also independent of the sample thickness due to the unique crystal structure and weak interlayer coupling (see **Fig. 3f**). The strength of $\sigma$ peak is of the same order as first distinct peak of conventional FE BaTiO$_3$ [27]. More importantly, the layer-independent $\sigma$ makes NbOCl$_2$ a promising material in the field of energy conversion beyond other 2D materials, because the NLO responses like $\sigma$ in TMDs will be suppressed in even-layer thickness [15], which is unsuitable in realistic energy applications.

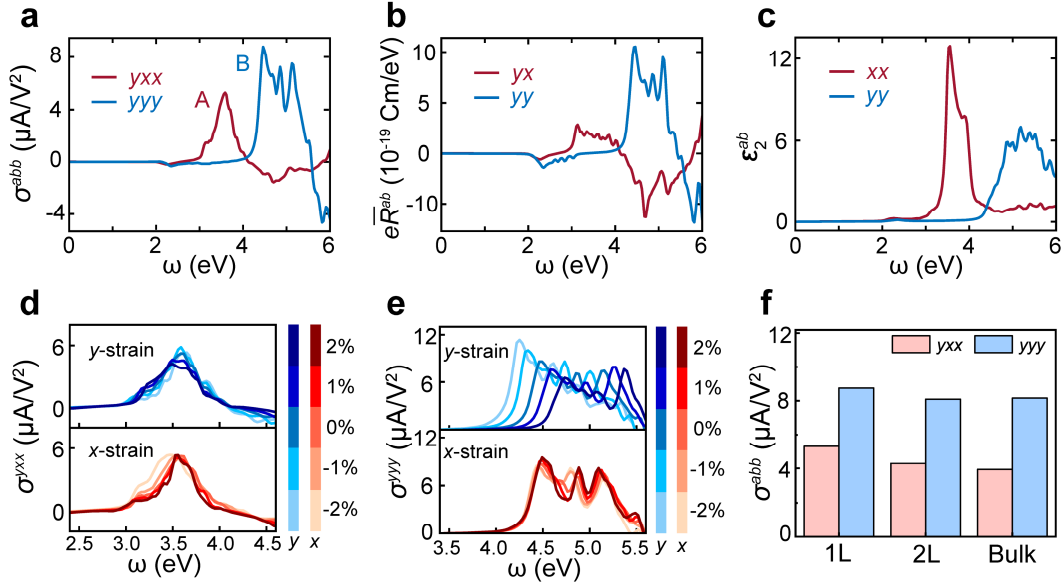

**Fig. 3 Shift Current in NbOCl₂. a-c** Frequency-dependent shift current susceptibility tensor $\sigma^{abb}$, integrated shift vector $e\bar{R}^{ab}$, and optical absorption spectrum $\varepsilon_2^{ab}$ induced by the *y*-polarized and *x*-polarized incident light, respectively. **d, e** Frequency-dependent $\sigma^{yxx}$ and $\sigma^{yyy}$ under uniaxial in-plane strains along *x* or *y* direction, respectively. **f** Comparison of the peak A in $\sigma^{yxx}$ and peak B in $\sigma^{yyy}$ with different layer thicknesses in NbOCl₂.

**External-field-tunable NLO Responses in NbOCl₂.** It is important to develop effective ways to control the NLO responses in NbO$X_2$. As proposed in **Fig. 1b**, the Nb atoms shifted along ***b*** direction can create a degree of freedom, which in principle may also generate different metastable AFE phases with the inversion symmetry that can effectively turn off the second-order NLO responses.

Whether any AFE phase can exist in NbO$X_2$ is still unknown in the experiments. In order to explore the possibility of realizing AFE phase in NbOCl₂ along with the non-centrosymmetric/centrosymmetric control, we have synthesized the single crystals and performed X-ray structure analysis of NbOCl₂ as a function of external pressure and temperature. Compared with other structural characterizations, we find that the single-crystal X-ray diffraction data can provide more accurate information on polymorphic structures of NbOCl₂. **Figure 4a** shows the high-pressure experimental setup. At the ambient condition, the bulk NbOCl₂ exists the FE phase. As shown in **Fig. 4b**, the off-center displacements of Nb atoms along *b* direction cause them to reside on one side of the Cl-atom-formed plane (labeled as $M_{Cl}$ plane, **Fig. 4b**), generating a non-centrosymmetric polarized structure. Interestingly, when we increase the pressure up to 5.7 GPa at room temperature or change the temperature from room temperature to a higher temperature of 500 K at ambient pressure, the collected single-crystal X-ray diffraction reveals that $NbOCl_2$ undergoes a novel FE-to-AFE phase transition (see Supplemental **Fig. S5** for detailed X-ray diffraction patterns). The structural analysis shows that the AFE structures maintain the vdW layered structure at 5.7 GPa or/and 500 K. Comparing the AFE and FE structures, we find that the pressure or temperature mainly affects the Nb-O bond length but has fewer influences on the Nb-Cl bonds and interlayer coupling. As shown in **Fig. 4c**, the nearly linear Nb-O-Nb-O chains along *b* direction are almost constant, while the Nb-O bonds change with the pressure or temperature (1.78 Å and 2.05 Å at 5.7 GPa and 1.781 Å and 2.119 Å at 500 K). More importantly, the off-center displacements of Nb atoms along *b* direction redistribute the Nb atoms on both sides of the $M_{Cl}$ plane, forming a centrosymmetric non-polarized structure with the twofold rotational symmetry retaining (space group of C2/c). In this AFE phase, every Nb pair composed of two adjacent Nb atoms shifts along different directions, separating them with the distance *l* along *b* direction (0.373 Å at 5.7 GPa and 0.437 Å at 500 K). Interestingly, this AFE phase has never been predicted. The detailed crystallographic information and structural refinement parameters measured in our experiments are listed in Supplemental **Tables S1-S10**. We emphasize that the subtle structure changes cannot be well detected by powder X-ray diffraction patterns.

The change of non-centrosymmetric structure to the inversion-symmetry-invariant structure also has important impacts on all the second-order NLO responses. Taking the SHG as an example, as shown in **Fig. 4d**, the normalized SHG intensity of single-crystal $NbOCl_2$ with transmission light $\lambda_{2\omega}$ of 532 nm is collected as a function of pressure under room temperature. Interestingly, the SHG intensity experiences three stages: (1) When the pressure is below ~3 GPa ($P < P_1$, stage I), the normalized SHG intensity remains constant, indicating the survival of FE phase at low pressure (**Fig. 4b**). (2) When the pressure is between ~3 and ~5.7 GPa ($P_1 < P < P_2$, stage II), the AFE phase starts to appear and mix with the FE phase. The higher the pressure, the larger the ratio of AFE phase over FE phase. Therefore, the SHG intensity continues to decrease. (3) When the applied pressure is above ~5.7 GPa ($P > P_2$, stage III), the FE sample is fully converted to the AFE one (**Fig. 4c**), finishing the phase transition. Eventually, the SHG intensity vanishes [43]. Even with the further increase of pressure, the SHG signal does not show up. In a similar way, we have performed the SHG measurements with the temperatures varying from 300 K to 600 K at ambient pressure. As shown in **Fig. 4e**, the SHG intensity also undergoes three similar stages, i.e., from FE phase (stage I) to FE-AFE mixing phase (stage II) and to AFE phase (stage III). The normalized SHG intensity starts to decrease at $T_1$~400 K and eventually reaches zero at $T_2$~450 K. When we release the pressure or cool down the temperature, the AFE phase becomes unstable turning back to the FE one. Besides SHG, it is believed that this FE-to-AFE phase transition can also continuously tune the shift current responses from maximum to zero in three novel stages.

We note that the homogeneous structures with the existing in-plane AFE order are relatively rare, which so far are mainly found in inorganic or organic-inorganic perovskites [44,45], molecular crystals [46] and 2D vdW $In_2Se_3$ [47]. Therefore, it is quite valuable for the observation of intralayer AFE order in a brand-new material system $NbOCl_2$. To reveal the origin of this reversible FE-to-AFE phase transition in $NbOCl_2$ under temperature or pressure, we have constructed the minimum energy path and obtained energy from the initial FE configuration and intermediate AFE configuration and final FE configuration using the DFT-based nudged elastic band (NEB) method. As shown in **Fig. 4f**, the total energy of the AFE metastable state is only 2 meV/u.c. higher than that of the FE ground state. There is an energy barrier of ~155 meV between FE and AFE phases, i.e., the phase transition can occur under a small pressure or a thermal perturbation. In addition, our calculations also reveal that the total energy of the observed AFE state is ~24 meV/u.c. lower than that of the predicted AFE state [48].

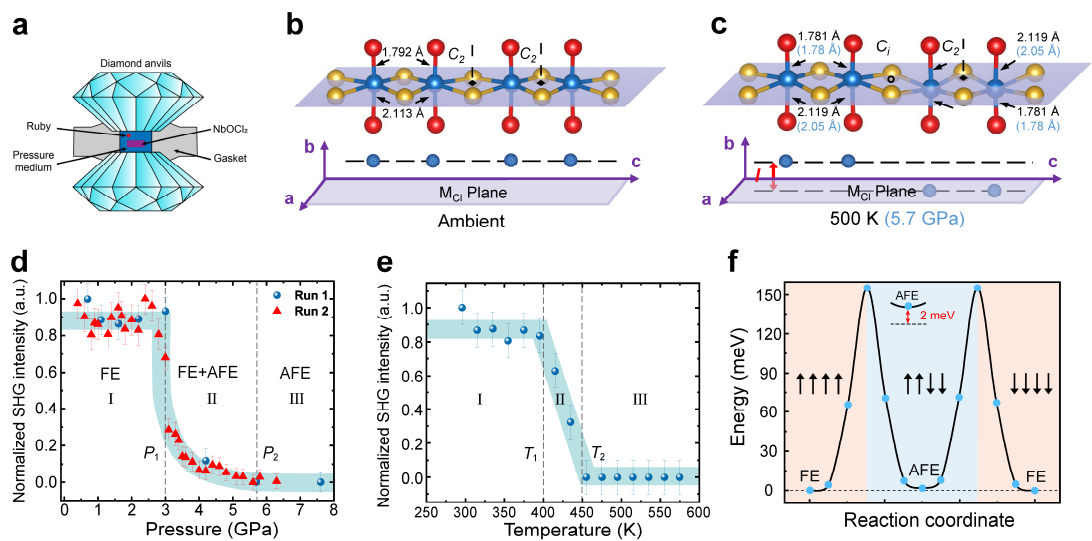

**Fig. 4 Reversible FE-to-AFE Phase Transition in NbOCl$_2$. a** Experimental high pressure setup, pressure is generated by two opposite diamond anvils. A gasket with a central hole is placed between two anvils to provide a chamber for samples and pressure transmitting medium. **b** Crystal structure of the FE NbOCl$_2$ at 300 K or/and ambient pressure determined by our experiments. Schematic diagram in the lower panel shows the Nb atoms on one side of the M$_{Cl}$ plane. **c** Crystal structure of the AFE NbOCl$_2$ at 500 K (black) or 5.7 GPa (blue). Schematic diagram in the lower panel shows the Nb atoms resided on both sides of the M$_{Cl}$ plane, with a distance *l*. Corresponding structural parameters in **b** and **c** are marked (see details in Supplemental **Table S1 and S6**). **d** Normalized SHG intensity as a function of external pressure (two different runs are performed under pressure). **e** Normalized SHG intensity as a function of temperature. **f** Total energy of NbOCl$_2$ as a function of generalized coordinates. Black arrows represent the direction of the Nb atomic displacements.

## Discussion

It is interesting to understand the evolution of NLO properties as a function of $X$ in NbO$X_2$. As shown in **Fig. 5a**, the calculated SHG strengths for peak B in these three different NbO$X_2$ are nearly identical (see full SHG spectra in Supplemental **Fig. S6**), consistent with the experimental observations [18,19]. For the optical transition near the peak of $\chi^{(2)}_{yyy}$, we find that double-photons resonance in NbOI$_2$ and NbOBr$_2$ comes from $X\ p \rightarrow$ Nb $d$, while O also contributes in NbOCl$_2$ ($X\ p +$ O $p \rightarrow$ Nb $d$) (see Supplemental **Fig. S7**), resulting in an order of NbOCl$_2$ > NbOI$_2$ > NbOBr$_2$. Besides, the similar band nesting effect [**Fig. 2e**] exists in all these NbO$X_2$ but in different zones of $k$ space (see Supplemental **Fig. S7**). Together with other virtual energy terms, these two main factors integrate over the entire Brillouin zone accidentally result in the strength of SHG less sensitive to the different $X$.

Different from the SHG, the $\sigma$ is noticeably sensitive to the $X$ (see full $\sigma$ spectra in Supplemental **Fig. S6**). As shown in **Fig. 5b**, the peak A of $\sigma^{yxx}$ [marked in **Fig. 3a**] increases with changing $X$ from Cl to I, but the peak B of $\sigma^{yyy}$ shows an opposite trend. To understand this unexpected phenomenon, we have calculated the $k$-resolved optical transition and $k$-resolved shift vector for these two peaks. For the $x$-polarized incident light, the optical transition is enhanced owing to the increase of $X$ orbitals in energy from Cl to I (see **Fig. 5c**, *upper* panel). Since the peak A of $\sigma^{yxx}$ is largely dominated by the $\varepsilon^{xx}_2$ (**Fig. 3c**), the enhancement of optical transition can increase values of peak A. On the other hand, for the $y$-polarized incident light, the shift vector is gradually weakening when changing $X$ from Cl to I (see **Fig. 5c**, *bottom* panel). Because the $\sigma^{yyy}$ is mainly determined by the shift vector (**Fig. 3b**), the weakening of shift vector results in the decrease of peak B.

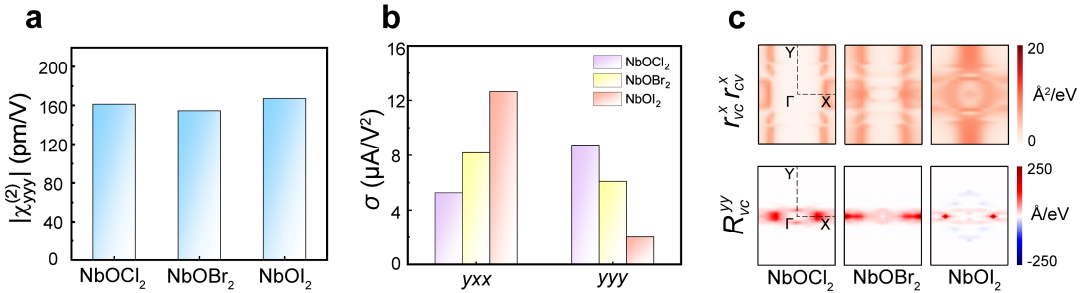

**Fig. 5 NLO Properties of Monolayer NbO$X_2$. a** Intensity of peak B in NbO$X_2$ with $X$ changing from Cl to I. Feature of peak B is marked in **Fig. 2b**. **b** Intensities of peak A in $\sigma^{yxx}$ and peak B in $\sigma^{yyy}$ with $X$ changing from Cl to I in NbO$X_2$. Features of peaks A and B are marked in **Fig. 3a**. **c** ***k***-resolved optical transition rate for peak A (*upper* panel) and shift vector for peak B (*bottom* panel) for monolayer NbO$X_2$.

In summary, combining theoretical and experimental studies, we present a systematical understanding of the NLO properties in the new NLO family materials NbO$X_2$. Theoretically, we reveal the physical mechanism of unusual NLO effects in NbO$X_2$, which not only can explain the key experimental observations, but also can deepen our understanding on the origin of NLO responses in 2D FE materials. Experimentally, we provide a powerful approach to effectively manipulate the NLO properties in NbOCl$_2$ continuously from maximum to zero as a result of crystal symmetry control, opening great opportunities for manipulating NbO$X_2$-based optoelectronic applications from optical sensing/computing/switch to BPVE.

**Acknowledgments**

This work is supported by the NSFC (Grants Nos. 12088101 and 12174404), National Key Research and Development of China (Grant No. 2022YFA1402400), NSAF (Grant No. U2230402), and the National Science Fund for Distinguished Young Scholars (Grant No. T2225027). The calculations were done in Tianhe-JK cluster at CSRC. Some experiments are supported by the Synergic Extreme Condition User Facility.

**Authors contributions**

B.H. conceived and led the project. H.G. supervised the experiments. L.T.Y. performed all the calculations. X. J. developed the computational code for NLO calculations and helped with the calculations. Q.G. synthesized the samples. W.Z., D.J., and H.G. performed the single crystal analysis. D.H., J.D., Y.W., and H.G. performed the experimental SHG measurements. L.T.Y., Y.L., W. Z. H.G., and B.H. analyzed all the data. L.T.Y., Y.L., W. Z., H.G., and B.H. wrote the manuscript with input from all other authors.

**References (for main text)**

[1] Hsieh, D., McIver, J. W., Torchinsky, D. H., Gardner, D. R., Lee, Y. S. and Gedik, N. Nonlinear Optical Probe of Tunable Surface Electrons on a Topological Insulator. *Phys. Rev. Lett*. **106**, 057401 (2011).

[2] Li, Y., Rao, Y., Mak, K. F., You, Y., Wang, S., Dean, C. R., and Heinz, T. F. Probing Symmetry Properties of Few-Layer $MoS_2$ and h-BN by Optical Second-Harmonic Generation. *Nano Lett.* **13**, 3329-3333 (2013).

[3] Sun, Z., Yi, Y., Song, T., Clark, G., Huang, B., Shan, Y., Wu, S., Huang, D., Gao, C., Chen, Z., McGuire, M., Cao, T., Xiao, D., Liu, W.-T., Yao, W., Xu, X. and Wu, S. Giant nonreciprocal second-harmonic generation from antiferromagnetic bilayer CrI3. *Nature* **572**, 497–501 (2019).

[4] Chu, H., Roh, C. J., Island, J. O., Li, C., Lee, S., Chen, J., Park, J.-G., Young, A. F. Lee, J. S., Hsieh, D. Linear Magnetoelectric Phase in Ultrathin $MnPS_3$ Probed by Optical Second Harmonic Generation. *Phys. Rev. Lett.* **124**, 27601 (2020).

[5] Van Aken, B., Rivera, J. P., Schmid, H. and Fiebig, M. Observation of ferrotoroidic domains. *Nature* **449**, 702–705 (2007).

[6] Leuthold, J., Koos, C., and Freude, W. Nonlinear silicon photonics. *Nat. Photon.* **4**, 535–544 (2010).

[7] Yao, K., Finney, N. R., Zhang, J., Moore, S. L., Xian, L., Tancogne-Dejean, N., Liu, F., Ardelean, J., Xu, X., Halbertal, D., Watanabe, K., Taniguchi, T., Ochoa, H., Asenjo-Garcia, A., Zhu, X., Basov, D. N., Rubio, A., Dean, C. R., Hone, J. and Schuck, P. J. Enhanced tunable second harmonic generation from twistable interfaces and vertical superlattices in boron nitride homostructures. *Sci. Adv.* **7**, eabe8691 (2021).

[8] Autere, A., Jussila, H., Dai, Y., Wang, Y., Lipsanen, H. and Sun, Z. Nonlinear Optics with 2D Layered Materials. *Adv. Mater.* **30**, 1705963 (2018).

[9] Li, M., Ling, J., He, Y., Javid, U. A., Xue, S. and Lin, Q. Lithium niobate photonic-crystal electro-optic modulator. *Nat. Commun.* **11**, 4123 (2020).

[10] Almeida, V., Barrios, C., Panepucci, R. and Lipson, M. All-optical control of light on a silicon chip. *Nature* **431**, 1081–1084 (2004).

[11] Cook, A., M. Fregoso, B., de Juan, F., Coh, S. and Moore, J. Design principles for shift current photovoltaics. *Nat. Commun.* **8**, 14176 (2017).

[12] Osterhoudt, G. B., Diebel, L. K., Gray, M. J., Yang, X., Stanco, J., Huang, X., Shen, B., Ni, N., Moll, P. J. W., Ran, Y. and Burch, K. S. Colossal mid-infrared bulk photovoltaic effect in a type-I Weyl semimetal. *Nat. Mater.* **18**, 471–475 (2019).

[13] Shockley, W., Queisser, H. J. Detailed balance limit of efficiency of pn junction solar cells. *J. Appl. Phys.* **32**, 510 (1961).

[14] Yin, X., Ye, Z., Chenet, D. A., Ye, Y., O'Brien, K., Hone, J. C. and Zhang, X. Edge nonlinear optics on a MoS$_2$ atomic monolayer. *Science* **344**, 488-490 (2014).

[15] Zhang, Y. J., Ideue, T., Onga, M., Qin, F., Suzuki, R., Zak, A., Tenne, R., Smet, J. H. and Iwasa, Y. Enhanced intrinsic photovoltaic effect in tungsten disulfide nanotubes. *Nature* **570**, 349–353 (2019).

[16] Kim, C.-J., Brown, L., Graham, M. W., Hovden, R., Havener, R. W., McEuen, P. L., Muller, D. A., Park, J. Stacking order dependent second harmonic generation and topological defects in *h*-BN bilayers. *Nano Lett.* **13**, 5660–5665 (2013).

[17] Wang, H. and Qian, X. Giant Optical Second Harmonic Generation in Two-Dimensional Multiferroics. *Nano Lett.* **17**, 5027-5034 (2017).

[18] Abdelwahab, I., Tilmann, B., Wu, Y., Giovanni, D., Verzhbitskiy, I., Zhu, M., Berté, R., Xuan, F., Menezes, L. de S., Eda, G., Sum, T. C., Quek, S. Y., Maier, S. A. and Loh, K. P. Giant second-harmonic generation in ferroelectric NbOI$_2$, *Nat. Photon.* **16**, 644-650 (2022).

[19] Guo, Q., Qi, X. Z., Zhang, L., Gao, M., Hu, S., Zhou, W., Zang, W., Zhao, X., Wang, J., Yan, B., Xu, M., Wu, Y.-K., Eda, G., Xiao, Z., Yang, S. A., Gou, H., Feng, Y. P., Guo, G.-C., Zhou, W., Ren, X.-F., Qiu, C.-W., Pennycook, S. J. and Wee, A. T. S.

Ultrathin quantum light source with van der Waals NbOCl$_2$ crystal. *Nature* **613**, 53-59 (2023).

[20] Dean, J. J. and van Driel, H. M. Second harmonic generation from graphene and graphitic films. *Appl. Phys. Lett.* **95**, 261910 (2009).

[21] McIver, J. W., Hsieh, D., Drapcho, S. G., Torchinsky, D. H., Gardner, D. R., Lee, Y. S. and N. Gedik. Theoretical and experimental study of second harmonic generation from the surface of the topological insulator Bi$_2$Se$_3$. *Phys. Rev. B* **86**, 035327 (2012).

[22] Rijnsdorp, J., Jellinek, F. The crystal structure of niobium oxide diiodide NbOI$_2$. *J. Less-Common Met*. **61**, 79–82 (1978).

[23] Hillebrecht, H., Schmidt, P.J., Rotter, H.W., Thiele, G., Zönnchen, P., Bengel, H., Cantow, H.-J., Magonov, S.N. and Whangbo, M.-H. Structural and scanning microscopy studies of layered compounds MCl$_3$ (M = Mo, Ru, Cr) and MOCl$_2$ (M = V, Nb, Mo, Ru, Os). *J. Alloys Compd*. **246**, 70–79 (1997).

[24] Beck, J. and Kusterer, C. Crystal structure of NbOBr2. *Z. Anorg. Allg. Chem.* **632**, 2193–2194 (2006).

[25] Tan, H., Li, M., Liu, H., Liu, Z., Li, Y. and Duan, W. Two-dimensional ferromagnetic-ferroelectric multiferroics in violation of the d$^0$ rule. *Phys. Rev. B* **99**, 195434 (2019).

[26] Zhang, Y., Lin, L.-F., Moreo, A. and Dagotto, E. Orbital-selective Peierls phase in the metallic dimerized chain MoOCl$_2$. *Phys. Rev. B* **104**, L060102 (2021).

[27] Young, S. M. and Rappe, A. M. First Principles Calculation of the Shift Current Photovoltaic Effect in Ferroelectrics. *Phys. Rev. Lett.* **109**, 116601 (2012).

[28] Yang, D., Wu, J., Zhou, B. T., Liang, J., Ideue, T., Siu, T., Awan, K. M., Watanabe, K., Taniguchi, T., Iwasa, Y., Franz, M. and Ye, Z. Spontaneous-polarization-induced photovoltaic effect in rhombohedrally stacked MoS$_2$. *Nat. Photon.* **16**, 469–474 (2022).

[29] Jahn, H. A. and Teller, E. Stability of Polyatomic Molecules in Degenerate Electronic States. I. Orbital Degeneracy. *Proc. R. Soc. London, Ser. A*, **161**, 220−235 (1937).

[30] Sipe, J. E. and Shkrebtii, A. I. Second-order optical response in semiconductors. *Phys. Rev. B* **61**, 5337-5352 (2000).

[31] Dresselhaus, M. S., Dresselhaus, G. and Jorio, A. *Group theory: application to the physics of condensed matter* (Springer, 2008)

[32] Ichiro, S., Takashi, K., Ayako, K., Masayuki, S. and Ryoichi I., Absolute scale of second-order nonlinear-optical coefficients. *J. Opt. Soc. Am. B* **14**, 2268-2294 (1997).

[33] Carvalho, A., Ribeiro, R. M. and Castro Neto, A. H. Band nesting and the optical response of two-dimensional semiconducting transition metal dichalcogenides. *Phys. Rev. B* **88**, 115205 (2013).

[34] Mennel, L., Smejkal,V., Linhart, L., Burgdörfer, J., Libisch, F. and Mueller, T. Band Nesting in Two-Dimensional Crystals: An Exceptionally Sensitive Probe of Strain. *Nano Lett.* **20**, 4242-4248 (2020).

[35] Jiang, X., Kang, L. and Huang, B. Role of interlayer coupling in second harmonic generation in bilayer transition metal dichalcogenides. *Phys. Rev. B* **105**, 045415 (2022).

[36] Zeng, H., Liu, G.-B., Dai, J., Yan, Y., Zhu, B., He, R., Xie, L., Xu, S., Chen, X., Yao, W. and Cui, X. Optical signature of symmetry variations and spin-valley coupling in atomically thin tungsten dichalcogenides. *Sci Rep* **3**, 1608 (2013).

[37] Tian, Z., Guo, C., Zhao, M., Li, R. and Xue, J. Two-Dimensional SnS: A Phosphorene Analogue with Strong In-Plane Electronic Anisotropy. *ACS Nano* **11**, 2219-2226 (2017).

[38] Trolle, M. L., Seifert, G. and Pedersen, T. G. Theory of excitonic second-harmonic generation in monolayer $MoS_2$. *Phys. Rev. B* **89**, 235410 (2014).

[39] Lihm, J. M. and Park, C. H. Wannier Function Perturbation Theory: Localized Representation and Interpolation of Wave Function Perturbation. *Phys. Rev. X* **11**, 041053 (2021).

[40] Rangel, T., Fregoso, B. M., Mendoza, B. S., Morimoto, T., Moore, J. E. and Neaton, J. B. Large Bulk Photovoltaic Effect and Spontaneous Polarization of Single-Layer Monochalcogenides. *Phys. Rev. Lett.* **119**, 067402 (2017).

[41] Yan, X., Li, P., Wei, S. H. and Huang, B. Universal Theory and Basic Rules of Strain-Dependent Doping Behaviors in Semiconductors. *Chin. Phys. Lett*. **38** 087103 (2021).

[42] Zhang, C., Pi, H., Zhou, L., Li, S., Zhou, J., Du, A., Weng, H. Switchable topological phase transition and nonlinear optical properties in a $ReC_2H$ monolayer. *Phys. Rev. B* **105**, 245108 (2022).

[43] Jiang, D., Song, H., Wen, T., Jiang, Z., Li, C., Liu, K., Yang, W., Huang, H., Wang, Y. Pressure-Driven Two-Step Second-Harmonic-Generation Switching in $BiOIO_3$. *Angew. Chem. Int. Ed.* **61**, e202116656 (2022).

[44] Mischenko, A. S., Zhang, Q., Scott, J. F., Whatmore, R. W., Mathur, N. D. Giant Electrocaloric Effect in Thin-Film $PbZr_{0.95}Ti_{0.05}O_3$. *Science* **311**, 1270-1271 (2006).

[45] Li, P. F., Liao, W. Q., Tang, Y. Y., Ye, H. Y., Zhang, Y. and Xiong, R. G. Unprecedented Ferroelectric–Antiferroelectric–Paraelectric Phase Transitions Discovered in an Organic–Inorganic Hybrid Perovskite. *J. Am. Chem. Soc.* **139**, 8752-8757 (2017).

[46] Lasave, J., Koval, S., Dalal, N. S., Migoni, R. L. Origin of Antiferroelectricity in $NH_4H_2PO_4$ from First Principles. *Phys. Rev. Lett*. **98**, 267601 (2007).

[47] Xu, C., Chen, Y., Cai, X., Meingast, A., Guo, X., Wang, F., Lin, Z., Lo, T. W., Maunders, C., Lazar, S., Wang, N., Lei, D., Chai, Y., Zhai, T., Luo, X. and Zhu, Y. Two-Dimensional Antiferroelectricity in Nanostripe-Ordered $In_2Se_3$. *Phys. Rev. Lett.* **125**, 047601 (2020).

[48] Jia, Y., Zhao, M., Gou, G., Zeng, X. C. and Li, J. Niobium oxide dihalides $NbOX_2$: a new family of two-dimensional van der Waals layered materials with intrinsic ferroelectricity and antiferroelectricity. *Nanoscale Horiz.* **4**, 1113-1123 (2019).

## Methods

**Density Functional Theory Calculations**. Density functional theory (DFT) [49,50] calculations were performed using the Vienna *ab initio* simulation package (VASP) [51]. Here, the projector-augmented wave method (PAW) [52] was employed to treat the core electrons. The 520 eV energy cutoff for the plane-wave basis and the Γ-centered 5 × 8 × 1 $k$-point mesh were used to ensure the well convergence for the calculated results. A 20 Å vacuum was applied to avoid the interaction introduced by the periodic boundary conditions. For all the calculations, the Perdew-Burke-Ernzerhof (PBE) functional within the framework of generalized-gradient approximation [53] was used to treat the exchange-correlation term in the Kohn-Sham equation. For the band structure calculations, the Heyd-Scuseria-Ernzerhof (HSE06) hybrid functional [54] was adopted, which can correct the PBE-calculated bandgap closer to the experimentally measured one. Crystal structures were fully relaxed until the Hellmann-Feynman forces < 0.015eV/Å for each atom. The criterion for total energy convergence was set to be 1×10$^{-6}$ eV to ensure the accuracy of the results.

**Nonlinear Optical Responses Calculations**. The NLO susceptibility tensors were calculated using our homemade package NOPSS, which can accurately calculate the SHG response [35,55], shift/injection currents, and photo-induced nonlinear spin current [56]. The SHG susceptibility tensor $\chi^{(2)}_{abc}(\omega)$ can be expressed as [57,58]

$$\chi^{(2)}_{abc}(-2\omega,\omega,\omega) = \chi^{abc}_{ter}(-2\omega,\omega,\omega) + \chi^{abc}_{tra}(-2\omega,\omega,\omega) \ (3),$$

where the first term represents the SHG susceptibility tensor contributed by purely interband effects and the second term originates from the mixing contribution of intraband and interband. Here, the $\chi^{abc}_{ter}$ and $\chi^{abc}_{tra}$ were calculated by

$$\chi^{abc}_{ter}(-2\omega,\omega,\omega) = \frac{e^3}{\hbar^2\Omega}\sum_{nml,\mathbf{k}}\frac{r^a_{nm}\{r^b_{ml}r^c_{ln}\}}{(\omega_{ln}-\omega_{ml})} \times \left[\frac{2f_{nm}}{\omega_{mn}-2\omega} + \frac{f_{ln}}{\omega_{ln}-\omega} + \frac{f_{ml}}{\omega_{ml}-\omega}\right] (4),$$

and

$$\chi^{abc}_{tra}(-2\omega,\omega,\omega) = \frac{i}{2}\frac{e^3}{\hbar^2\Omega}\sum_{nm,\mathbf{k}}f_{nm}\left[\frac{2}{\omega_{mn}(\omega_{mn}-2\omega)}r^a_{nm}(r^b_{mn;c}+r^c_{mn;b}) + \frac{1}{\omega_{mn}(\omega_{mn}-\omega)}(r^a_{nm;c}r^b_{mn}+r^a_{nm;b}r^c_{mn}) + \frac{1}{\omega^2_{mn}}\left(\frac{1}{\omega_{mn}-\omega}-\frac{4}{\omega_{mn}-2\omega}\right)r^a_{nm}(r^b_{mn}\Delta^c_{mn}+r^c_{mn}\Delta^b_{mn}) - \frac{1}{2\omega_{mn}(\omega_{mn}-\omega)}(r^b_{nm;a}r^c_{mn}+r^c_{nm;a}r^b_{mn})\right] (5).$$

The shift current tensor $\sigma_2$ was calculated using the formula [30]

$$\sigma^{abc}(0;\omega,-\omega) = -\frac{i\pi e^3}{2\hbar^2}\int \frac{d\mathbf{k}}{8\pi^3}\sum_{nm} f_{nm}(r^b_{mn}r^c_{nm;a} + r^c_{mn}r^b_{nm;a})\delta(\omega_{mn}-\omega) \quad (6).$$

When the light is limited to polarize linearly in the $b$ direction, it can also be expressed more concisely as

$$\sigma^{abb}(0;\omega,-\omega) = \frac{\pi e^3}{\hbar^2}\int \frac{d\mathbf{k}}{8\pi^3}\sum_{n,m} f_{nm} R^{ab}_{nm} |r^b_{nm}|^2 \delta(\omega_{mn}-\omega) \quad (7),$$

where the shift vector is defined by

$$R^{ab}_{nm}(\mathbf{k}) = \frac{\partial \phi^b_{nm}(\mathbf{k})}{\partial k^a} - \xi^a_{nn}(\mathbf{k}) + \xi^a_{mm}(\mathbf{k}) \quad (8).$$

To explore the origin of $\sigma$, the integrated shift vector $e\bar{R}^{ab}$ and linear optical absorption spectrum $\varepsilon^{ab}_2$ were employed, which are given by [40]

$$e\bar{R}^{ab} = e\Omega \int \frac{d\mathbf{k}}{8\pi^3}\sum_{n,m} f_{nm} R^{ab}_{nm}\delta(\omega_{nm}-\omega) \quad (9),$$

and

$$\varepsilon^{ab}_2(-\omega;\omega) = \frac{\pi e^2}{\hbar}\int \frac{d\mathbf{k}}{8\pi^3}\sum_{n,m} f_{nm} r^a_{nm} r^b_{mn}\delta(\omega_{mn}-\omega) \quad (10).$$

Among Eqs. (3)-(10), the connection $\xi_{nm} = \frac{i(2\pi)^3}{\Omega}\int d\mathbf{r}\, u^*_n(\mathbf{k},\mathbf{r})\nabla_{\mathbf{k}} u_m(\mathbf{k},\mathbf{r})$. If $n \neq m$, $r_{nm} = \xi_{nm}$, otherwise $r_{nm}=0$, where $\Omega$ is the volume of the unit cell, $u_m(\mathbf{k},\mathbf{r})$ is the periodic part of Bloch wavefunction. Superscripts $\{n, m, l\}$ and $\{a, b, c\}$ represent the band indices and Cartesian indices, respectively. $\{r^b_{ml} r^c_{ln}\} = \frac{1}{2}(r^b_{ml} r^c_{ln} + r^c_{ml} r^b_{ln})$. $f_{nm} = f_n - f_m$, where $f_n(f_m)$ is the Fermi distribution function. $\hbar\omega_{nm} = \hbar\omega_n - \hbar\omega_m$, where $\hbar\omega_n(\hbar\omega_m)$ is energy. $r^b_{nm;a}$ is the generalized derivative of the matrix element $r^b_{nm}$, which can be written as $r^b_{nm;a} = \frac{r^a_{nm}\Delta^b_{mn} + r^b_{nm}\Delta^a_{mn}}{\omega_{nm}} + \frac{i}{\omega_{nm}} \times \sum_l \omega_{lm} r^a_{nl} r^b_{lm} - \omega_{nl} r^b_{nl} r^a_{lm}$. $\Delta^b_{mn} = v^b_{mm} - v^b_{nn}$ represents the difference of the electron group velocities. $\phi^b_{nm}$ is the phase of $r^b_{nm} = |r^b_{nm}|e^{i\phi^b_{nm}}$ and the shift vector follows the relationship $R^{ab}_{nm} = \frac{\text{Im } r^b_{nm} r^b_{mn;a}}{r^b_{nm} r^b_{mn}}$.

Here we employ the scissor operator from HSE06 functional calculations to get an accurate band gap. The dense of $k$-point mesh and the number of bands were carefully examined for convergence, and the 19×32×1 k-point mesh and 120 electronic bands were finally chosen to perform calculations (see Supplemental, **Fig. S8**). For all calculations, 200 grids were linearly interpolated in the energy range of 0~6 eV to make the curve smooth, and the small imaginary smearing factor appeared in the susceptibility tensors expression was set to be $\eta = 0.05$ eV, a common value used for many 2D NLO material calculations [17,35]. In addition, to compare the NLO effects of systems with different thicknesses, the effective susceptibility was employed, which was defined as $\chi_{eff}^{NLO} = \frac{\chi_{bulk}^{NLO} * L_s}{\frac{L_a}{2}}$, where $L_s$ is the thickness of the slab model and $L_a$ is the lattice constant of the bulk NbO$X_2$.

**Experimental Setup.** The details of crystal growth were described in our previous article [19]. High-pressure experiments were carried out in a Mao-Bell type diamond anvil cell (DAC) with 300 μm culet anvils and steel gaskets. Neon was loaded as the pressure-transmitting medium. Ruby fluorescence is used for the pressure measurement [59]. The DAC was mounted and centered on a Bruker D8 Venture four-circle diffractometer according to Dawson et al. [60]. Single crystal X-ray diffraction data were collected at 5.7 GPa and 293 (2) K with multilayer monochromator Mo Kα radiation (λ = 0.71073 Å). Data were collected in ω-scans in eight settings of 2θ and ϕ with a frame and step size of 60 s and 0.5°, respectively. Data collection strategy was based on that described by Dawson et al. [60]. The sample reflections were harvested manually by using the SMART code. Image masks, to avoid integrating the signal from detector regions obscured by DAC, were created using the program ECLIPSE [61]. For high temperature single crystal x-ray diffraction data, a black block-shaped single crystal of NbOCl$_2$ with dimensions of 0.063×0.043×0.039 mm$^3$ was selected and fixed on the top of a thin glass fiber. X-ray diffraction data were collected increasing the temperature to 500(2) K using an Oxford Cryosystems Plus low-temperature device and were measured on a Bruker D8 Venture four-circle diffractometer with multilayer monochromator Mo Kα radiation (λ = 0.71073 Å). Data integration and oblique correction was performed with the software package SAINT [62]. Absorption effects were corrected using the Multi-Scan method (SADABS) [63]. Structures were solved by dual space methods (SHELXT) and refined by full-matrix least-squares on $F^2$ (SHELXL) [64] using the graphical user interface ShelXle [65]. All atoms were refined anisotropically for the room-pressure structure and the high-pressure structures.

**References (for Method section)**

[49] Hohenberg, P. and Kohn, W. Inhomogeneous Electron Gas. *Phys. Rev.* **136**, B864-B871 (1964).

[50] Kohn, W. and Sham, L. J. Self-Consistent Equations Including Exchange and Correlation Effects. *Phys. Rev.* **140**, A1133-A1138 (1965).

[51] Furthmüller, J. and Kresse, G. Efficient iterative schemes for ab initio total-energy calculations using a plane-wave basis set. *Phys. Rev. B* **54**, 11169-11186 (1996).

[52] Blöchl, P. E. Projector augmented-wave method. *Phys. Rev. B* **50**, 17953-17979 (1994).

[53] Perdew, J. P., Burke, K. and Ernzerhof, M. Generalized Gradient Approximation Made Simple. *Phys. Rev. Lett.* **77**, 3865-3868 (1996).

[54] Heyd, J., Scuseria, G. E. and Ernzerhof, M. Hybrid functionals based on a screened Coulomb potential. *J. Chem. Phys.* **118**, 8207 (2003).

[55] Li, P., Jiang, X., Huang, M., Kang, L., Chen, S., Gali, A. and Huang, B. Defect engineering of second-harmonic generation in nonlinear optical semiconductors. *Cell Rep. Phys. Sci.* **3**, 101111 (2022).

[56] Jiang, X., Ye, L., Wu, X., Kang, L. and Huang, B. Role of large Rashba spin-orbit coupling in second-order nonlinear optical effects of polar $BiB_3O_6$. *Phys. Rev. B* **106**, 195126 (2022).

[57] Rashkeev, S. N., Lambrecht, W. R. L., Segall, B. Efficient ab initio method for the calculation of frequency-dependent second-order optical response in semiconductors. *Phys. Rev. B* **57**, 3905-3919 (1998).

[58] Wang, C., Liu, X., Kang, L., Gu, B.-L., Xu, Y. and Duan, W. First-principles calculation of nonlinear optical responses by Wannier interpolation. *Phys. Rev. B* **96**, 115147 (2017).

[59] Mao, H. K., Xu, J. A. and Bell, P. M. Calibration of the ruby pressure gauge to 800 kbar under quasi-hydrostatic conditions. *J. Geophys. Res.: Solid Earth* **91**, 4673 (1986).

[60] Dawson, A., Allan, D. R., Parsons, S. and Ruf, M. Use of a CCD diffractometer in crystal structure determinations at high pressure. *J. Appl. Cryst.* **37**, 410-416 (2004).

[61] Parsons, S. ECLIPSE--Program for masking high-pressure diffraction images and conversion between CCD image formats. Software (2010).

[62] Bruker AXS Inc, APEX3 Ver. 2016.1-0. Madison, Wisconsin, USA, (2012).

[63] G.M. Sheldrick, SADABS, Version 2008/1, Bruker AXS Inc., 2008.

[64] Sheldrick, G. M. A short history of SHELX. *Acta Crystallogr. A* **64**, 112–122 (2008).

[65] Hübschle, C.B., Sheldrick, G.M., and Dittrich, B. ShelXle: A Qt graphical user interface for SHELXL. *J. Appl. Cryst.* **44**, 1281-1284 (2011).